# Intergranular and intragranular properties of superconducting multifilamentary Bi–2223 / Ag tapes : comparison of electrical transport and magnetic measurement methods.

**J.-F. Fagnard**[a,c], **Ph.Vanderbemden**[a], **A. Genon**[a], **R. Cloots**[b] and **M. Ausloos**[c]

SUPRAS, University of Liège, Sart-Tilman, B-4000 Liège, Belgium.
(a) Department of Electrical Engineering and Computer Science, Montefiore Institute B28
(b) Chemistry Institute B6
(c) Physics Institute B5
e-mail : fagnard@montefiore.ulg.ac.be

**Abstract**
Several experimental techniques have been used in order to characterize the properties of multifilamentary Bi-2223 / Ag tapes. Pristine samples were investigated by electrical resistivity, current-voltage characteristics and DC magnetic moment measurements. Much emphasis is placed on comparing transport (direct) and magnetic (indirect) methods for determining the critical current density as well as the irreversibility line and resolving usual lacks of consistency due to the difference in measurement techniques and data analysis. The effect of an applied magnetic field, with various strengths and directions, is also studied and discussed. Next, the same combination of experiments was performed on bent tapes in order to bring out relevant information regarding the *intergranular* coupling. A modified Brandt model taking into account different types of defects within the superconducting filaments is proposed to reconciliate magnetic and transport data.

## 1. Introduction

The critical current density $J_C$ represents the key factor determining the usefulness of High-Temperature Superconductors (HTS) for engineering applications. The granular microstructure of HTS requires to distinguish between *intragranular* critical currents flowing within individual grains and *intergranular* critical currents flowing across grain boundaries. The connectivity between grains seems not to be very detrimental for Bi-based materials [1] which are up to now the best candidates for applications involving extended lengths of superconductor. Nowadays, $Bi_2Sr_2Ca_2Cu_3O_{10}$ (hereafter called Bi-2223) tapes and cables of hundreds meters are available from several manufacturers [2-4]. Considerable progress has also been achieved in processing tapes containing several filaments embedded in a silver matrix ("multifilamentary tapes"). The main advantages of multifilamentary tapes as compared to monofilamentary ones are (i) a reduction of AC losses [5-8], (ii) an improved current distribution uniformity [8,9] and (iii) better mechanical properties [1,8].

In spite of promising current carrying properties over long distances, the major disadvantage of Bi-based materials is their low irreversibility field $H_{irr}$, which lies around 0.1 T at 77 K [10]. The direct consequence is a sharp decrease of the critical current density with the magnetic field at relatively "high" temperatures ($T \sim 77$ K) [11,12]. Related problem is due to the thermally activated motion of flux lines (flux creep), yielding some dissipation for $J < J_c$, i.e. non-linear $E$-$J$ relations ($E \sim J^n$) [13,14]. Limitations to tape performances are also expected to be due to macroscopic defects such as cracks, cross-section irregularities or secondary non-superconducting phases. Indeed, the transport supercurrent has been shown to be quite inhomogeneous [15,16]: the critical current density varies along the length as well as the thickness of the tape, the highest values of $J_c$ being located near the edges of the superconducting filaments, close to the Ag sheath. The conductor inhomogeneities are expected to combine to flux creep in reducing the above $n$-value, but both contributions cannot be distinctly sorted out from the usual experiments [17]. An additional characteristic of Bi-2223 materials is their high anisotropy, leading to a severe dependence of the critical current density with respect to the magnetic field direction [13,18-20].

The exact reasons for current flow degradation in superconducting tapes represent thus an intensive research area. All the physical mechanisms listed above do have an impact on the $J_c(H)$ behaviour. However, the crucial issue is to clarify which are predominant and how $J_c(H)$ could be improved. At present, it has been experimentally observed that different $J_c$ limiting mechanisms generally operate in different magnetic field ranges. More precisely, four regimes have been pointed out [19,21] :
(i) for very small applied fields, the self-field dominates and $J_c$ is almost independent of $H$,
(ii) at low field, $J_c$ exhibits a power-law dependence $J_c \sim H^{-n}$ characteristic of a progressive breaking of weak-links,
(iii) at high field, $J_c(H)$ is controlled by flux motion within the grains themselves and displays an exponential form,
(iv) once the irreversibility field is exceeded, $J_c$ drops to zero.

Understanding the electrical and magnetic behaviour of superconducting tapes necessitates reliable characterization techniques. Basically, investigations can be achieved either by *transport* or *magnetic* measurements. Each technique brings out specific information, and their comparison has been shown to be helpful in elucidating the current limitation processes in HTS tapes [22]. The two main parameters relevant to applications, the critical current density $J_c$ and the irreversibility field $H_{irr}$ or irreversibility line (*IL*), can be extracted and values reconciled from various experiments, but keeping the following considerations in mind.

First let us consider the critical current density determination. In the case of electrical transport measurements, the *V-I* relationship imposes to choose an arbitrary criterion (e.g. 1 µV / cm) for defining the appearance of a finite voltage across the sample and determining the value of the critical current $I_c$. From this value, *the critical current density $J_c$ can be deduced by dividing $I_c$ by the supercurrent cross-section*. However, the latter could be much less than the total superconductor cross-section since the carrying transport current mainly percolates through the less resisitive path [21]. Another problem is related to current redistribution between the superconducting core and the silver sheath, particularly near localized defects; these effects were shown to produce additional linear sections in the *V-I* curves [23,24]. The situation becomes more intricate in the case of multifilamentary tapes for which the outer filaments may tend to screen the inner ones and carry most of the current [8]. By using Hall probe scanning, some authors [25] have also shown that the magnetic coupling between filaments results in a current distribution corresponding to an "effective" individual filament.

Consider now the current density determination using magnetic methods, e. g. DC magnetization measurements [26]. In DC magnetic experiments, the irreversible magnetization can be translated into $J_c$ using the Bean model [27]. However, great care should be taken on the following points : (i) both inter- and intra- granular currents may contribute to the magnetic moment [21,26,28-30], (ii) for $H \parallel c$, the flat geometry of the sample needs to incorporate appropriate models of flux penetration [31], (iii) for $H \parallel ab$, the shielding is due to currents flowing in the *ab*-plane *and* along the *c*-axis, thereby requiring to take the $J_c$ anisotropy into account [26], (iv) the lengthscale over which shielding currents flow (i.e. the average size of the current loops) can be less than the whole sample size [21,32].

Let us consider now the case of the irreversibility line (*IL*) determination. At this stage, it is worth mentioning that the *IL* is not a "true" well-defined line because different measurement techniques yield different values.

First, the *irreversibility field $H_{irr}$* can be determined from transport measurements using two methods. The first one is based on defining $H_{irr}$ as the field above which dissipation takes place and electrical resistance emerges from the noise level, or exceeds a given criterion. The second method defines $H_{irr}$ as the field at which the curvature of *E-J* (or *V-I*) characteristic on a log-log plot changes from negative to positive. Both techniques were shown to give similar $H_{irr}$ values provided that the tape is sufficiently homogeneous [9], which seems to be the case for multifilamentary tapes. Since both techniques are based on the observation of a given threshold voltage – or resistance (e.g. 0.4 µΩ in the present paper), the *resistive irreversibility line* represents thus a line of constant resistance located just *above* the "true" irreversibility line.

On the other hand, the *magnetic irreversibility line* is the field beyond which the difference between the upper and the lower branch of the hysteretic loop is below a non-zero threshold, and thus a non-zero $J_c$. This *magnetic* irreversibility line is thus expected to be located *below* the "true" irreversibility line in the technological diagram *H-T*. Experimentally, this inevitably corresponds to some small **D***m* depending on both the sample size and the resolution of the experimental technique used. This feature has to be taken into account for properly explaining magnetic measurement results.

Another relevant point is the role played by the electric field *E* in magnetic experiments, as highlighted by Campbell [33] and Caplin [34]. The motion of vortices – caused either by a time-varying applied magnetic field or by magnetic relaxation – leads to a non zero *dB/dt* whence, according to Faraday's law, to a small intrinsic electric field *E*. The consequence is that magnetic measurements usually explore much lower voltage regimes than transport measurements do. The criterion for defining $J_c$ is therefore dependent of the characterizing technique.

In our group, we have studied different superconductors in particular YBCO-based and Bi-2223 samples through electrical and magnetic properties characterization from intragrain and intergrain feature point of view as related to the chemical synthesis process. Several commercial measurements devices were used but, in many cases, specialized home-made ones had to be developed. This is the case for the critical current measurement device used here or a susceptometer [35-40].

The aim of the present work is to resolve such of these constraints following investigations of both the electrical and the magnetic properties of *multifilamentary* Bi-2223 / Ag tapes with various measurement techniques including electrical resistivity, *I-V* curves and DC magnetization. By comparing the transport and magnetic properties, we propose to describe several features characteristic of the tapes. This paper is organized as follows : in Section 2, the experimental details are briefly described. Section 3 deals with the characterization of the tapes by transport and magnetic properties as well as the link which can be established between both types of measurements. Finally, in Section 4, conclusions are drawn from the analysis given in the preceding sections.

## 2. Experimental techniques

### 2.1. Microstructure

The multifilamentary Bi-2223 / Ag tapes studied in this paper have been prepared with the power-in-tube (PIT) technique [8] and are commercialized by Nordic Superconductor Technology (NST) [2]. The tape cross-section is (3.4 x 0.22) mm², and the overall superconducting cross-section is equal to 0.25 mm². The filament microstructure consists of small platelet-like grains of typical dimensions (1 x 10-20 x 10-20) µm³. The Bi-2223 compound is textured such that grains align within 5° to 10° with their *c*-direction perpendicular to the tape largest side.

### 2.2. Transport measurements

Electrical resistance as a function of temperature *R(T)* and *V(I)* curves were measured by the conventional 4-points method, using a Quantum Design Physical Property Measurement System (PPMS). This commercial device allows us to apply high magnetic fields with various field orientations *q* with respect to the sample, but is limited to currents up to 2 A. A home-made pulsed-current set-up was used to perform high critical current measurements up to 40 A, with applied magnetic fields up to 0.7 T. Both systems are thus complementary because of the decrease of the critical current with respect to the magnetic field.

The measurements at high magnetic field and low current are performed with the PPMS and the low magnetic field range is covered by the home-made device. For all transport measurements, the critical current was extracted using a 1 µV / cm threshold. At such voltage levels, the effect of a parallel current channel through the silver matrix is estimated to be negligible.

### 2.3. Magnetic measurements

DC magnetic measurements at several temperatures and magnetic fields were performed in the Physical Property Measurement System using an extraction method. This method allows for a typical magnetic moment resolution of about $10^{-7}$ Am². During any given M-H loop recording, the field was changed always using a constant sweep rate, $\mu_0\, dH/dt \sim 1$ mT / s.

In order to estimate the relative importance of *intergranular* and *intragranular* currents contributions $m_J$ and $m_G$, the grain boundaries were next damaged by severely bending the tape up to a 1 mm radius, as proposed by Müller et al. [30]. The tape properties were then measured after straightening. In this case, the intergranular part of the response is expected to be almost suppressed and only the intragranular component $m_G$ to remain.

## 3. Results and discussion

### 3.1. Transport critical current

The field dependence of the tape critical current is shown in Fig. 1. At $T = 77$ K and self-field, the critical current is 40 A, corresponding to a critical current density of about $1.6\ 10^4$ A/cm².

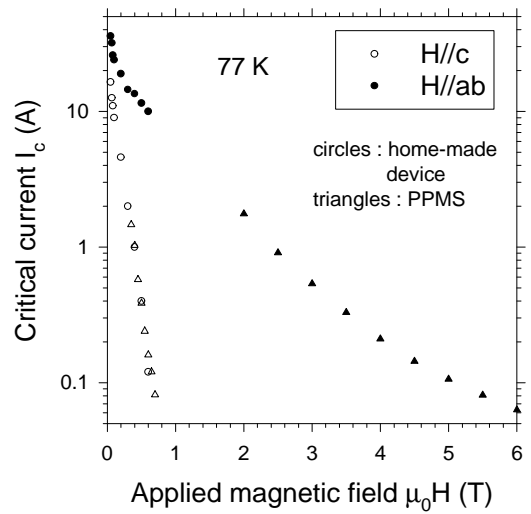

**Figure 1.** Critical current of Bi-2223/Ag tapes as a function of magnetic field parallel to the c-axis or parallel to the ab-plane as aligned in the tape at 77 K.

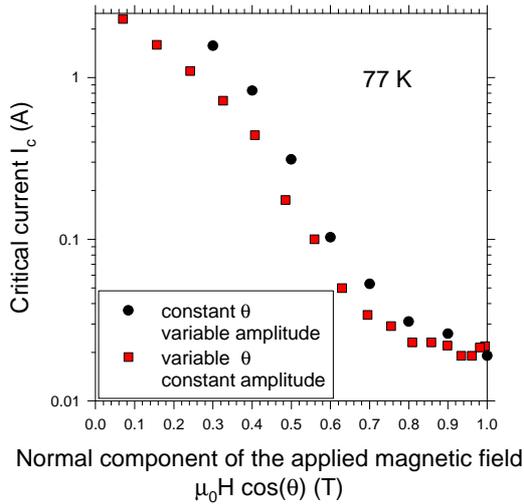

**Figure 2.** Critical current versus the normal component of the applied magnetic field at 77 K.

Note that this value represents a lower bound for the true $J_c$ since (i) the current carrying cross-section might be smaller than the total cross-section of the superconductor and (ii) $J_c$ varies along the thickness of the tape. However, for a multifilamentary tape containing small filaments as here, this latter effect is expected to be not very pronounced. At $\mu_0H = 0.5$ T, the critical current drops to 12 A for $H \parallel ab$ and to 0.4 A when $H \parallel c$ (Fig.1).

The strong anisotropy of the critical current is outlined in Fig. 2 which compares the $I_c(H)$ data for $H \parallel c$ scaled to the field component in the c direction ($H \cos \theta$) with respect to the angular dependance $I_c(\theta)$ for $\mu_0H = 1$ T. As can be seen, both data are similar to each other in a large range of the magnetic field normal component. The anisotropic behaviour depicted in Fig. 2 is in good agreement with reports in the literature [13,19] and underlines the deleterious influence of the normal component of the field on the Bi-2223 tape critical current.

### 3.2. Magnetic moment measurements

Figure 3 shows the magnetic moment $m(H)$ curves measured at $T = 77$ K for $H \parallel c$ and $H \parallel ab$ (with different scales) on a piece of the original tape before (white symbols) and after (black symbols) the bending process described above. We remark that the overall magnetization is about one order of magnitude larger when $H \parallel c$ compared to the case $H \parallel ab$. As expected [28,29], the magnetization of the bent tape is always smaller than the magnetization of the virgin tape due to the disappearance of *a part of* the intergranular currents contribution to the magnetic moment. An interesting issue to be addressed is to know whether nearly *all* intergranular junctions have disappeared or whether the currents still percolate through strongly connected islands, as suggested in ref. [13]. Considering that *all* grain boundaries are deteriorated by the bending is questionable. Therefore we consider that we measure, in the bent sample, a small contribution of intergranular currents in remaining intact grain clusters. It has however to be noticed that the difference between bent and unbent tapes is rather small (less than 20%) when $H \parallel ab$ (Fig. 3), compared to 50% when $H \parallel c$, therefore confirming that the bending mostly affects the links in the *ab*-plane.

However, it should be noticed that, despite the high degree of texturing in the studied tape, some of the grains are expected not to be perfectly aligned with each other. These imperfections are likely to have an impact when the magnetic field is parallel to the tape surface (what is referred as "$H \parallel ab$"). In such a case, some grains will experience a small magnetic field component parallel to the *c*-axis which might reduce the expected signal after the tape was bent.

The measurements on the bent tapes will be used in Sect. 3.5 for extracting the magnetic and transport intergranular critical current densities.

a) 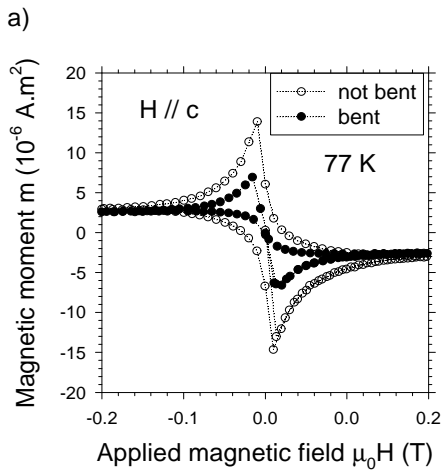 b) 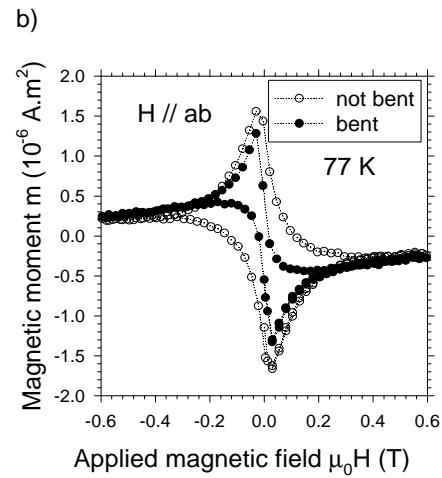

**Figure 3.** Magnetic moment of bent and unbent tapes for an applied magnetic field a) parallel to the *c*-axis or b) parallel to the *ab*-plane as aligned in the tape at 77 K.

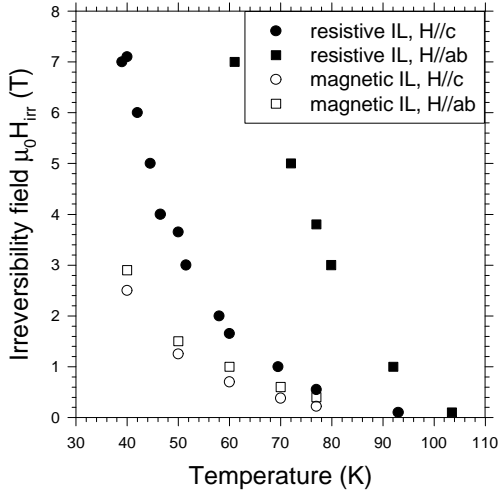

**Figure 4.** Resistive and magnetic irreversibility lines for an applied magnetic field parallel to *c*-axis or parallel to the *ab*-plane.

*3.3. Irreversibility lines*

Magnetic moment data recorded at several temperatures were used in order to determine the *magnetic* irreversibility line from the $H$ value where the upper and the lower branches of the hysteresis loop converge. The *resistive* irreversibility line (*IL*) was obtained by determining the field at which the electrical resistance exceeds the noise level [9-11], defined here as 0.4 $\mu\Omega$. The data obtained by both experimental methods are summarized in Fig. 4. As can be seen, the magnetic *IL*'s determined for both orientations (white symbols) lie always well below the resistive *IL*'s (black symbols).

In the present case, the striking difference between both techniques can be justified in terms of the criterions used. For magnetic measurements, the magnetic field variation (lying here in the range 1-10 mT/s) induces an electric field whose value depends on the size of the current loops. For $H \parallel c$, if we consider that the lengthscale to be looked at is the average filament half-width (100 μm), the corresponding electric field is about $10^{-8}$ V/m. Such a value lies well within the range $10^{-11}$-$10^{-7}$ V/m reported by *Caplin et al.* [34] and is roughly two orders of magnitude below the electric field observed in the transport method (3 $10^{-6}$ V/m). Magnetic measurements carried out for various sweep rates $dH/dt$ have shown that an electric field increase of one order of magnitude corresponds to a 40 % irreversibility field augmentation. This suggests that magnetic measurements using very large magnetic sweep rates, not achievable through our measuring set-up but giving rise to electric fields comparable to those associated with transport experiments, mately double the present magnetic irreversibility field value. At T = 40 K, this would bring the magnetic $\mu_0 H_{irr}$ from 2.5 T up to 5 T, which is still lower than the measured resistive $\mu_0 H_{irr}$ (7 T). This suggests, *in turn*, that the average length scale over which shielding currents flow is somewhat smaller than the average filament half width. For $H \parallel ab$, this length scale is very small, on the order of the filament half-thickness (10 μm), corresponding to a very small electric field value ($10^{-9}$ V/m) This explains the much smaller pronounced disagreement between the resistive and magnetic *IL* values for $H \parallel ab$ than for $H \parallel c$.

Irreversibility line measurements combining resistive and magnetic methods for different field orientations are not often discussed in the literature [35]. Kiuchi *et al.* [12] shows similar results to ours and also point out the difference between the irreversibility lines for $H \parallel c$ and $H \parallel ab$ but only by using the magnetic method. For $H \parallel c$, the measurements presented here agree with those obtained by Grasso [36]. From our results we can conclude that considerable caution has to be taken for a proper determination of the irreversibility line but we have drawn how to resolve disagreements.

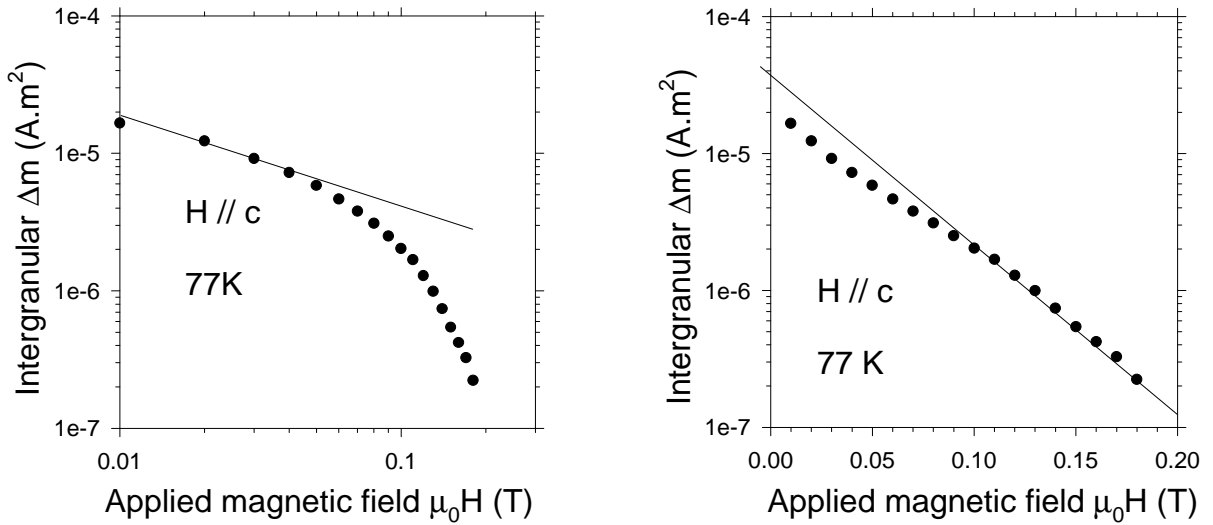

**Figure 5.** Intergranular *Δm* in log-log and log-linear plots.

*3.4. Intragranular and intergranular behaviours*

The behaviour of the irreversible magnetic moment depends on the magnetic field amplitude. The existence of two regimes, revealed by measurements performed on bent and unbent tapes, has been established by numerous authors [19,21,22]. The observed magnetic moment on intact tapes is due to the shielding currents flowing over the whole sample (typically 0.3 x 4 mm² if $H \parallel c$) and those circulating on smaller dimensions (grains or clusters of grains). Even after some strong bending, some clusters of grains are expected to be still connected, but these clusters are likely to be much less than the bending radius (1 mm), thus much less than the sample size. For the intact tape, the low field behaviour can be understood as an intergranular dissipation in the weak-links, leading to the power-like law typical in percolation. This dependance is shown in Fig. 5 on a log-log plot. On this figure is presented the same data using a semi-log scale. It can be seen that, for fields exceeding 0.1 T, the behaviour is exponential, suggesting that the flux motion inside the grains is responsible for the $J_c$ degradation, as explained in ref [37].

In the case of the bent tape (Fig. 6), the decrease is seen to be almost wholly. The fact that the low field power-law regime has disappeared shows that the intergranular current contribution to the magnetic moment is nearly totally suppressed. At high fields, the characteristic decay field $\mu_0 H_g$ of the $\Delta m = \Delta m_0 \exp(-\mu_0 H / \mu_0 H_g)$ law is the same at given temperature for the bent and the unbent tapes (Fig. 7). This result strongly confirms that the high field behaviour is dominated by the intragranular component. This also gives us confidence, *a posteriori*, that a large majority of the weak links has been destroyed by bending.

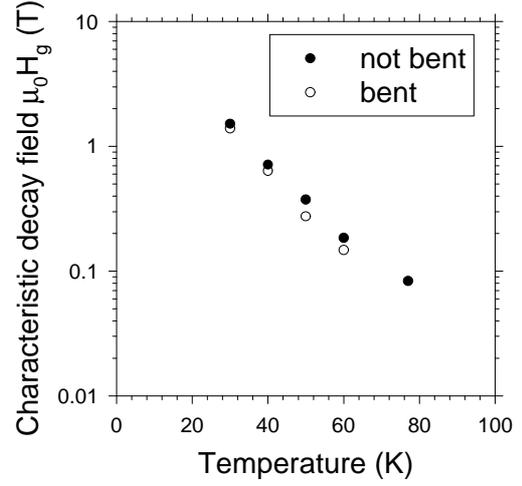

**Figure 7.** Characteristic decay field of the exponential $J_c(B)$ behaviour versus temperature for bent and unbent tapes.

*3.5. Reconciling transport and magnetic measurements differences*

The use of magnetic data for extracting the critical current density $J_c$ requires the knowledge of the relevant current flow geometry [22]. Transport measurements only give access to the current density flowing in the *ab*-planes. Therefore, in order to probe $J_c^{ab}$ through magnetic measurements, the magnetic field has to be applied parallel to the *c*-axis. In such a geometry, the filaments demagnetization factor must be taken into account and we use the flux penetration model in thin strips developed by Brandt [38]. According to this model, the relationship between the critical current and the magnetic moment is given by

$$m_{\downarrow\uparrow} = \pm \mu a^2 L H_c \left[ \tanh \frac{H_m}{H_c} + 2 \tanh \left( \frac{\pm H - H_m}{2 H_c} \right) \right], \quad (1)$$

where $m$ is the magnetic moment, $H$ is the applied field, $H_m$ its maximum value, $L$, $a$ and $d$ denote respectively the length, the half-width and the thickness of the tape. $H_c$ is a characteristic field related to $J_c$ by $H_c = J_c d / \mu$. The symbols $\downarrow$ and $\uparrow$ denote the branches of the hysteresis loop for a decreasing and an increasing applied magnetic field respectively.

In this model, the tape is supposed to be homogeneous, infinitely long and the current density independent of the local magnetic induction. Moreover, the condition $d << a$ must be satisfied. In our case, we assume that the total tape magnetic response is given by the sum of the magnetic moment of each filament, for which we apply the above relationship.

In order to probe the intergranular magnetic moment, data measured on original and bent samples were subtracted. Figure 8 shows the difference between the upper and the lower branches of the resulting intergranular hysteresis loop. On the same figure, the values calculated using Brandt model with a field dependent critical current density are plotted for comparison. Unexpectedly, the best critical current dependence law is $J_c = J_0 \exp(-H/H_0)$ where the characteristic decay field $H_0$. The intergranular critical current density at zero field, $J_0$, obtained by fitting the irreversible intergranular moment curve, is $8 \cdot 10^3$ A/cm². At 0.1 T, this value is $8.6 \cdot 10^2$ A/cm². It should be noticed that, in our case, the use of the 2D Bean model neglecting demagnetization effects applied to a sample of rectangular cross-section give similar results, i.e. $J_c \sim 10^3$ A/cm² at 0.1 T, in spite of the fact that this model is valid for infinite thickness sample. For $\mu_0 H = 0.1$ T $\parallel c$, the intergranular $J_c$ given by the transport pulses method is $3.6 \cdot 10^3$ A/cm² which is 3 to 4 times higher than the value given by the magnetic measurements. We consider now several mechanisms which could explain the disagreement. The ratio between the $J_c$ value determined by magnetic measurements ($J_c^M$) and the $J_c$ value obtained by the transport measurements ($J_c^T$) will be denoted by $q$. From the above results, its value is 0.24 at 0.1 T perpendicular to the tape plane (Table Ia-b).

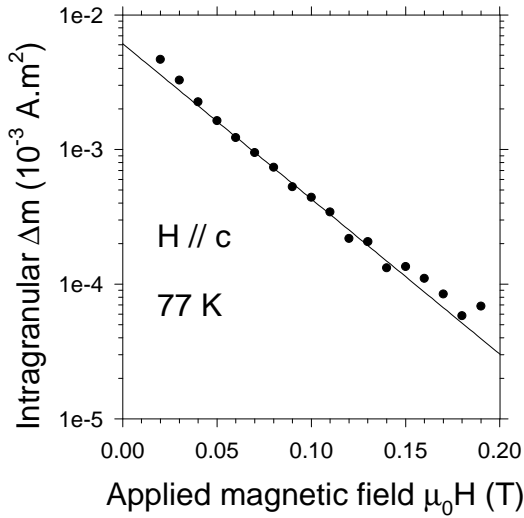

**Figure 6.** Intragranular $\Delta m$ in log-linear plot.

First of all, it should be emphasized that the magnetic moment obtained by subtracting data measured on original and bent samples can be reduced which respect to the true intergranular moment if some intergranular coupling remains in the bent tape. However, it is insufficient to explain a factor of 4 for the $J_c$'s determined by both methods.

The difference between $J_c^M$ and $J_c^T$ could be due to the different electric fields values involved in magnetic and transport experiments. However, as mentioned above, the magnetic measurements should yield a $J_c$ lower than that obtained from transport measurements because the used criterion in the former case is more severe.

Another way to explain the difference is to consider that the true geometry of the filaments is not strictly the one assumed above. First, the filaments do not have a perfect rectangular cross-section. Second, some defects can reduce the superconducting cross-section leading to underestimating $J_c^T$. The intergranular $J_c^M$ is also affected by the defects but not in the same way. These defects can prevent shielding currents from circulating over the whole filament. Therefore the critical current density determined from magnetic moment measurements $J_c^M$ is also underestimated. Taking all these considerations into account, we can consider several simple configurations in order to determine whether the existence of defects can lead to reconciliating the values given by transport and magnetic measurements.

In the following, we consider that the superconducting region is characterized by a homogeneous $J_c$ and we can study the expected ratio between $J_c^M$ and $J_c^T$ while varying the size of the non superconducting defect. The parameters $a$, $d$ and $L$ have the same signification as for the original Brandt model. We examine two types of "ideal" defects.

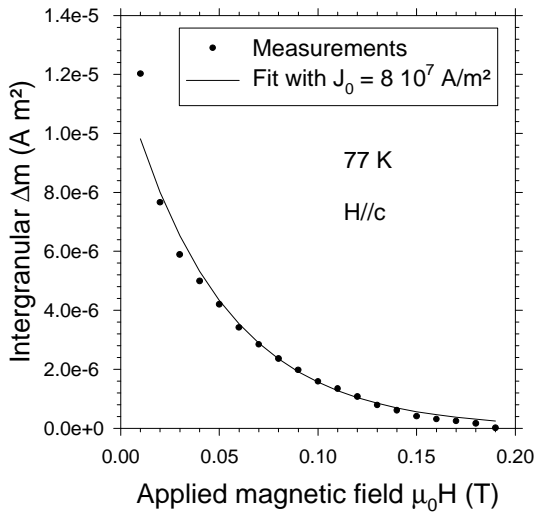

Figure 8. Comparison between the measurements of the intergranular $\Delta m$ at 77K with H//c and the $\Delta m$ fitted by combining the Brandt theory and a critical current density exponential law.

| a) Initial filament | b) **_Measurements at 0.1 T (H//c)_** |
|---|---|
| 2a, L, L >> 2a | Figure 1 : $m = m^{meas} = 7.9 \cdot 10^{-7}$ Am² <br> → $J_c^M = 8.67 \cdot 10^2$ A/cm² (Brandt) <br> → $J_c^M = 1.07 \cdot 10^3$ A/cm² (Bean) <br> Table I : $J_c^T = J_c^{meas} = 3.6 \cdot 10^3$ A/cm² <br> ⇒ $q = J_c^M/J_c^T = 0.24$ (Brandt) <br> $= 0.3$ (Bean) |
| c) Filament with longitudinal defect | d) **_Brandt model_** (infinite tape) |
| 2a, L, x | x = 0 : q = 0.53 <br> x = 2a/5 : q = 0.64 <br> x = 2a/3 : q = 0.77 <br> x = a : q = 1.05 |
| e) Filament with transverse defects | f) **_Bean model_** (rectangular sample) |
| 2a, 1, 2, N-1, x <br> N-1 defects, N segments | (3D plot of q vs x/2a and N) |

**Table I.** Critical current densities $J_c^M$ and $J_c^T$ at 0.1 T perpendicular to the tape.
    a, b : Geometry and results for the initial filaments.
    c, d : Geometry and results for filaments with longitudinal defect.
    e, f : Geometry and results for filaments with transverse defects.

The first one is a longitudinal defect caused by assuming that the central part of the filament is not superconducting (Table Ic). This assumption arises from the fact that the critical current density is not homogeneous along the cross-section of the filament, the $J_c$ value being higher near the silver matrix than near the filament centre [39]. In magnetic moment measurements, we consider that shielding current loops are confined within each region limited by these longitudinal defects.

The introduction of a longitudinal defect (Table Ic) leads to the modification of the parameter $a$ in Eq. (1). This parameter appears only as a prefactor in this expression, therefore it is possible to obtain the magnetic moment correction for several defect widths. We have obtained the $J_c^M$ value in each case by fitting the magnetic moment measurements with Eq. (1) taking into account the correction caused by the defect. The calculated ratio $q$ increases with the longitudinal defect width, $x$, and $q$ approaches 1 for $x \approx a$ (Table Id).

Table Ie shows transverse defects along the tape length. These defects define short segments where shielding current loops circulate more easily than in the whole sample. In this case, the Brandt theory, assuming that currents loops close at infinity, cannot be used. The tape must be considered as a cluster assembly from which we can sum up the magnetic moment of each $N$ tape segment. In order to calculate these magnetic moments, we use a simple Bean model applied to a rectangular sample with isotropic critical current density [40]. When the intact sample is fully penetrated, one has

$$m = \frac{J_c d(2a)^2}{4}\left(L - \frac{2a}{3}\right). \qquad (2)$$

While, if we consider $N$ isolated tape segments (as shown in Table Ie), one has

$$m = \frac{N J_c d(2a)^2}{4}\left(\frac{L}{N} - \frac{2a}{3}\right), \qquad (3)$$

with obvious geometrical relations.

The critical current density obtained by the transport method depends on the smallest effective cross-section through which the current is able to flow. The $q$ ratio varies when either the number or the width of the transverse defects changes (Table If). If the defect number increases, the $J_c$ values are such that $q$ is equal to 1. This occurs especially if the defects induce a tape segmentation without obstructing the current flow.

With such simple cases, we have proposed explanations of the magnetic and transport critical current densities difference by assuming the existence of defects that modify the theoretical values obtained by both techniques.

*3.6. Intragranular to intergranular critical current ratio*

It is of interest to use the data analysed above in order to estimate the intragranular critical current density from the intragranular hysteresis loop. In this case, the geometric effects can be excluded as being a relevant parameter affecting the results. The reason arises from the fact that the grains are aligned and stacked on the top of each other. Their self demagnetizing effect is thus reduced by the mutual magnetic shielding as compared to the single isolated grain case. We assume thus that the grain geometry consists of long cylinders with length $d$ (thickness of the filaments) with a 10 μm average radius $R_G$. The Bean model can be used and gives [29]:

$$J_{cG} = \frac{3\left(m_J^\uparrow - m_J^\downarrow\right)}{4 R_G a d L}. \qquad (4)$$

At 0.1 T, this model gives an intragranular critical current density equal to $6.6 \; 10^3$ A/cm². As mentioned above, the intergranular critical current density at 0.1 T from the Brandt model is $8.6 \; 10^2$ A/cm². The intergranular on intragranular $J_{cG}/J_{cJ}$ ratio at 77 K and 0.1 T is thus evaluated to lie around 5-10. For comparison, Müller et al [29] has calculated an intragranular critical current density $J_{cG}$ that is about a factor 100 larger than the intergranular critical current density $J_{cJ}$ within the Bean model at zero applied magnetic field and several temperatures for monofilamentary tapes. This ratio seems typical for 40 K and is reduced at higher temperatures.

**4. Conclusions**

The aim of this paper was to investigate basic properties of Bi-2223 / Ag tapes through transport and magnetic measurements and to compare the results.

First, a strong difference between magnetic and transport irreversibility fields has been pointed out. The main causes of the disagreement are the quite different intrinsic electric field and subsequent criteria involved in both techniques, combined with the fact that the true lengthscale over which shielding currents flow is smaller than the geometric one, defined by the superconducting cross-section perpendicular to the magnetic field. Such results show the great care to be taken about the experimental methods used for determining the irreversibility line.

The comparison between the *Dm* curves of bent and virgin tapes shows two different behaviours following the applied magnetic field amplitude. At low field, there is an intergranular regime while the high field regime is intragranular. The high field exponential law is the same whether the tape has been bent or not. This result confirms that, in this range of magnetic field, the current limitation is due to flux motion *in the grains* and not to weak-link dissipation.

Finally, the Brandt theory of the magnetic moment is used to compare the transport and the magnetic measurements. We have extended the model in order to explain the difference between the critical current densities determined by transport and magnetic measurements. We attribute this difference to the presence of various defects within the filaments. In addition, the intragranular on intergranular $J_{cG}/J_{cJ}$ ratio has been estimated to lie between 5 and 10 for a 0.1 T applied magnetic field.

## 7. Acknowledgements

This work is part of a contract with the Région Wallonne de Belgique (VESUVE). Ph. V. is recipient of a FNRS scientific research worker fellowship. We also would like to acknowledge Prof. H.W. Vanderschueren for allowing us to use MIEL laboratory facilities and V. Misson for his help on critical current measurements.